\documentclass[11.5pt]{article} 
\usepackage{wrapfig,amssymb,amsfonts,amsmath,here,color,hangcaption}
\usepackage{graphicx,here}

\renewcommand{\theequation}{\arabic{section}.\arabic{equation}} 
 
\makeatletter
\@addtoreset{equation}{section} 
\@addtoreset{footnote}{section} 
\makeatother 
 
\makeatletter 
\renewcommand\appendix{
\par
\setcounter{section}{0}%
\setcounter{subsection}{0}%
\gdef\thesection{Appendix \@Alph\c@section }
\renewcommand{\theequation}
{\Alph{section}.\arabic{equation}}
}
\makeatother 
 
\makeatletter
\def\eqnarray{ \stepcounter{equation} \let\@currentlabel=\theequation
\global\@eqnswtrue
\global\@eqcnt\z@
\tabskip\@centering
\let\\=\@eqncr
$$\halign to \displaywidth\bgroup\@eqnsel\hskip\@centering
$\displaystyle\tabskip\z@{##}$&\global\@eqcnt\@ne
\hfil$\displaystyle{{}##{}}$\hfil
&\global\@eqcnt\tw@$\displaystyle\tabskip\z@{##}$\hfil
\tabskip\@centering&\llap{##}\tabskip\z@\cr}
\makeatother

\makeatletter
\def\@arrayacol{\edef\@preamble{\@preamble \hskip .5\arraycolsep}}
\def\array{\let\@acol\@arrayacol \let\@classz\@arrayclassz
\let\@classiv\@arrayclassiv \let\\\@arraycr\def\@halignto{}\@tabarray}
\makeatother

\makeatletter
\newcounter{subeqncnt}
\def\thesubeqncnt{\alph{subeqncnt}}
\def\subequations{\begingroup%
\stepcounter{equation}\edef\@tempa{\theequation}%
\let\c@equation\c@subeqncnt\c@subeqncnt\z@
\edef\theequation{\@tempa\noexpand\thesubeqncnt}}

\makeatother

\newcommand{\captionfonts}{\small}
\makeatletter 
\long\def\@makecaption#1#2{%
\vskip\abovecaptionskip
\sbox\@tempboxa{{\captionfonts #1: #2}}%
\ifdim \wd\@tempboxa >\hsize
{\captionfonts #1: #2\par}
\else
\hbox to\hsize{\hfil\box\@tempboxa\hfil}%
\fi
\vskip\belowcaptionskip}
\makeatother 

\begin{document}

\setlength{\baselineskip}{7mm}
\begin{titlepage}
\begin{flushright}
\end{flushright}

\setlength{\baselineskip}{9mm}

\begin{center}

\vspace*{15mm}
{\LARGE Holographic Superconducting Quantum Interference Device}
 
\vspace{9mm}
{\Large{{Shingo Takeuchi}}$^{\dagger\,\ddagger}$
}

\setlength{\baselineskip}{0mm}

\vspace*{12mm}
$ \dagger$
{\large \it{Shanghai Jiao Tong University, Shanghai 200240, China}}\\
\vspace*{4mm}
$\ddagger$
{\large \it{\textcolor{black}{The Institute for Fundamental Study ``The Tah Poe Academia Institute''}}}\\
\vspace*{2mm}
{\large \it{\textcolor{black}{Naresuan University Phitsanulok 65000, Thailand}}}\\
\vspace*{5mm}
{\large \textcolor{black}{shingo(at)nu.ac.th}}
\end{center}

\vspace{7mm}

{\large
\begin{abstract}
We \textcolor{black}{present} a holographic model of the SQUID~(Superconducting QUantum Interference Device) 
in the \textcolor{black}{external magnetic field}. \textcolor{black}{The model of the gravitational theory considered in this paper} 
is the Einstein-Maxwell-complex scalar model on the four-dimensional Anti-de Sitter Schwarzschild black brane geometry, 
\textcolor{black}{where} one space direction is compacted into a circle and we arrange \textcolor{black}{the coefficient of the time components profile 
so that we can model the SQUID, where the profile plays the role of the chemical potential for the cooper pair.}
\end{abstract}
}

\end{titlepage}

\section{Introduction}
\label{Chap:Intro} 

In recent years, we have been studying the superconductor\textcolor{black}{s} 
using the potential applicability of anti-de Sitter~(AdS)/conformal field theory~(CFT) correspondence~\cite{Maldacena:1997re,Gubser:1998bc,Witten:1998qj}.  
The notable examples are Refs.\cite{Gubser:2008px,Hartnoll:2008vx,Hartnoll:2008kx}, Refs.\cite{Lee:2008xf,Liu:2009dm,Cubrovic:2009ye} and Ref.\cite{Nishioka:2009zj} 
which have provided the gravity duals for the superconductors, the (non-)Fermi liquids,  
and the superconductor/insulator transition at zero temperature, respectively. 
\textcolor{black}{
Further the  s-\,\,,\,p- and d-wave superconductors have been also thoroughly investigated. 
Although we cannot refer to each paper for these, for example see \cite{Cai:2015cya} for its synthetic report.}

One of the interesting topics associated with the superconductivity would be the Josephson junction~\cite{Josephson:1962zz}, 
and now there are many developing studies of the holographic version. The first paper was Ref.\cite{Horowitz:2011dz}, 
in which the superconductor-normal metal-superconductor~(SNS) Josephson junction was studied holographically.   
After that, toward Ref.\cite{Horowitz:2011dz}, 
the dimensional extension was discussed in Refs.\cite{Wang:2011rva,Siani:2011uj}, 
and generalized to the p-wave Josephson junction as discussed in Ref.\cite{Wang:2011ri}. 
Ref.\cite{Wang:2012yj} provided the gravity dual for a superconductor-insulator-superconductor~(SIS) Josephson junction. 
Ref.\cite{Kiritsis:2011zq} invented the holographic Josephson junction based on the designer multigravity~(namely multi-(super)gravity theories).
\textcolor{black}{Ref.\cite{Li:2014xia} constructed a Josephson junction in the non-relativistic case with a Lifshitz geometry as the dual gravity.}

The superconductors in the external magnetic field would be intriguing, and its holographic study has also been developing.  
The first papers are Refs.\cite{Hartnoll:2008kx, Nakano:2008xc, Albash:2008eh, Hartnoll:2007ai},  
and although we cannot refer to all the papers, the papers we have particularly checked in this paper are 
Refs.\cite{Domenech:2010nf, Montull:2011im, Salvio:2013ja, Montull:2012fy, Cai:2012vk} along with \textcolor{black}{those four papers.}

\begin{wrapfigure}{h}{60mm}
\begin{center}
\includegraphics[scale=0.60]{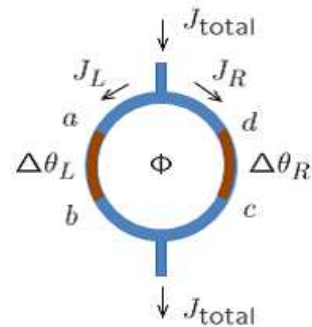}
\caption{
The SQUID {blue}{considered} in this paper. 
For the meaning of each character and the figure, see the body text.
}
\label{FigSQUID}
\end{center}
\end{wrapfigure}

This paper will be devoted to a holographic SQUID~(Superconducting QUantum Interference Device) 
consisting of two Josephson junctions with the external magnetic field shown in Fig.\ref{FigSQUID}. 
The significance in the holographic study of the Josephson junction and the SQUID would be the application to the case 
where the coupling constants become strong in a system occupied by two different parts.   
For instance, we take an issue with whether the two supercoductors can stick each other perfectly or not 
when they are put separately with some distance~($L$) in between some empty space without the mid materials. 
Since it is known that the free energy for such two supercoductors is a decreasing function 
when $L$ becomes smaller~\cite{JosephsonMaterial}, they move close to each other. 
Here we assume that no friction occurs when these supercoductors move. 
However, when $L$ becomes very small, the quantum effect becomes strong. As a result, it is difficult to conclude 
whether two supercoductors can stick to each other perfectly or not eventually. 
It is very interesting if this issue relates to the issue whether evaporating black holes vanish or remain eventually  
by regarding the empty space as the evaporating black holes and two superconductor parts as the space of the outside of the black holes.   
In such an issue, the strongly coupled analysis is very important, 
and the skill developed in the holographic Josephson junctions and the SQUID like ones in this paper could be a big help.

Lastly, we give a comment on the same type of the holographic SQUID in Ref.\cite{Cai:2013sua}. 
Originally, there are two supercurrents in the SQUID as represented as $J_L$ and $J_R$ in the left and the right sides of the SQUID in Fig.\ref{FigSQUID}.
On the other hand, in Ref.\cite{Cai:2013sua}, only one supercurrent is considered. 
Despite this, they measure the phase differences in the left and the right Josephson junctions, $\Delta \theta_L$ and $\Delta \theta_R$, 
where $\Delta \theta_L$ and $\Delta \theta_R$ are shown in Fig.\ref{FigSQUID}. 
However, their way to measure the phase differences from one current is not clear. 
This has motivated us to carry out this paper\footnote{\textcolor{black}{
This study had been performed with Ref.\cite{Cai:2013sua} as one study at first.  
However due to a scientific disagreement on this point, as a result of many discussion at the last stage of the work, 
finally I have come to publish this paper separating from Ref.\cite{Cai:2013sua}. 
For this reason the content of this paper is similar with Ref.\cite{Cai:2013sua}}}. 
In this paper, another way for the holograhic SQUID in the magnetic field is proposed, 
and we can confirm that we can obtain the result consistent with the original SQUID in the external magnetic field and the result in Ref.\cite{Cai:2013sua}.      
\newline

Regarding the organization of this paper;~ 
Section.\ref{Chap:Review_of_SQUID} is devoted to brief review of the SQUID we will consider in this paper:~the points in our holographic way to model the SQUID, 
which are how to take in the magnetic flux and how to take in the specific behavior of the supercurrent and the conclusion of this paper. 
In section.\ref{Chap:setup}, we give our gravity model and the equations of motion. 
Further, we mention about the ansatz.  
In section.\ref{analysis}, we show the results of our analysis in our holographic SQUID. 
In  \ref{app:1}, we explain the validity of a condition used in Section.\ref{Chap:Review_of_SQUID}.    
In \ref{app:2}, we list the numerical results  in section.\ref{analysis} explicitly.

\section{Brief review of SQUID, our holographic way to model it \textcolor{black}{and conclusion}}
\label{Chap:Review_of_SQUID}

We devote this section to a brief review of the SQUID 
we consider in this paper and the points in our holographic way to model it.

\subsection{Brief review of SQUID in this paper}
\label{SubChap:Review_of_SQUID}

The SQUID we consider in this paper is given by connecting two Josephson junctions 
in a circle and attaching two branches for the inflowing and outflowing supercurrents $J_{\rm total}$.  
We show it schematically in Fig.\ref{FigSQUID}.

In Fig.\ref{FigSQUID}, $\Phi$ represents the magnetic flux, and the blue and the brown parts represent the superconductor and the normal metal parts, respectively. 
\textcolor{black}{Then} the wave functions in the superconductor parts are condensed \textcolor{black}{in a} phase.   
As a result, the phase differences arise in the intervals of the normal metal parts $a-b$ and $c-d$,  
and $\Delta\theta_L$ and $\Delta\theta_R$ represent such phase differences in the intervals of the normal metal parts $a-b$ and $c-d$. 
As a result, the supercurrent is induced by the Josephson effect~\cite{Josephson:1962zz}, 
and  $J_L$ and $J_R$ represent such supercurrents flowing in the left and the right sides of the circuit of SQUID.  
These are known to be written with the sine function as 
\begin{eqnarray}
J_L = I_{cL} \sin \Delta \tilde{\theta}_L
\quad \textrm{and} \quad
J_R =  I_{cR} \sin \Delta \tilde{\theta}_R,
\end{eqnarray}
where $I_{cL}$ and $I_{cR}$ \textcolor{black}{are constants} meaning the maximal supercurrent 
and $\Delta \tilde{\theta}_L$ and $\Delta \tilde{\theta}_R$ \textcolor{black}{presents} the gauge invariant phase differences defined as
$\displaystyle
\Delta \tilde{\theta}_L \equiv -\frac{e^*}{\hbar} \int_a^b {\bf A} \cdot {\rm d{\bf l}}+ \Delta \theta_L
$ 
and 
$\displaystyle
\Delta \tilde{\theta}_R \equiv -\frac{e^*}{\hbar} \int_c^d {\bf A} \cdot {\rm d{\bf l}}+ \Delta \theta_R 
$, and we obtain this sine relation later in Fig.\ref{sine_relation}. 
\textcolor{black}{Here} $e^*$ \textcolor{black}{represents} the electric charge of a cooper pair forming the supercurrents, 
$d\bf l$ \textcolor{black}{represents} the line elements along the circuit of the SQUID and ${\bf A}$ \textcolor{black}{represents} the gauge field on the circuit, 
and `$a$', `$b$', `$c$', `$d$'  \textcolor{black}{are} the locations on the circuit in Fig.\ref{FigSQUID}.  
\textcolor{black}{Consequently}, the total amount of the inflowing supercurrent $J_{\rm total}$ can be written as
\begin{eqnarray}
\label{totalcurrent1}
J_{\rm total} &=& J_L - J_R
\nonumber\\
&=& 2 J_{c} 
\cos \left( \frac{\Delta \tilde{\theta}_L + \Delta \tilde{\theta}_R}{2} \right) 
\sin \left( \frac{\Delta \tilde{\theta}_L - \Delta \tilde{\theta}_R}{2} \right),
\end{eqnarray}
where we have assumed \textcolor{black}{that} $J_{cL}=J_{cR}\equiv J_{c}$ for simplicity 
and assigned the minus sign to $J_R$ taking into \textcolor{black}{account the} fact that 
we measure the phases in \textcolor{black}{an} anticlockwise direction in the circuit of Fig.\ref{FigSQUID}.

\textcolor{black}{Here} the contour integral of the infinitesimal variation of the phase $\nabla \theta$ 
along the circuit of the SQUID should be given by integral multiplication of $2\pi$ as 
$\displaystyle
2\pi n = \oint \nabla \theta \cdot  {\rm d{\bf l}}
$, 
where $n$ means some integer number and $\nabla \theta$ is known to be given in the condensed matter physics as 
$\nabla \theta = (m^* {\bf v}_s +e^* {\bf A}) /\hbar$~(${\bf v}_s$ and $m^*$ are the velocity and the mass of the cooper pairs, 
while $e^*$ has been defined above).   
Evaluating the right hand of it, it turns out that this contour integral can be written as 
\begin{eqnarray}\label{twopin2}
2\pi n = 2\pi \frac{\Phi}{\Phi_0} + \Delta \tilde{\theta}_L + \Delta \tilde{\theta}_R, 
\end{eqnarray}
where $\Phi_0 \equiv h/e^*$ and $\Phi$ means the magnetic flux penetrating the circuit of the SQUID,  
$\displaystyle 
\Phi = \int (\nabla \times {\bf A}) \cdot d {\bf S}$, 
where $d\bf S$ means the area element.  
\textcolor{black}{Here} to derive the above relation, ${\bf v}_s$ is taken to zero~(${\bf v}_s=0$) in the superconductor parts of the circuit of the SQUID
by assuming that the path of the integration goes through the center of the section of the circuit. 
We give a description \textcolor{black}{to validate} this ${\bf v}_s=0$ in \ref{app:1}.

\textcolor{black}{Then} using the relation (\ref{twopin2}),  eq.(\ref{totalcurrent1}) can be rewritten as
\begin{eqnarray}
\label{totalcurrent2}
J_{\rm total} &=& J_{\rm max}(\Phi) \sin \left( \Delta \tilde{\theta}_L + \pi \frac{\Phi}{\Phi_0} \right)
\quad \textrm{with} \quad J_{\rm max}(\Phi) \equiv 2 J_{c} \cos \left( \pi \frac{ \Phi}{\Phi_0} \right), 
\end{eqnarray}
where the above has been obtained with $n=0$ in eq.(\ref{twopin2}). 
The behavior of $J_{\rm max}(\Phi)$ against the magnetic flux $\Phi$ is one of the specific behavior of SQUID, 
and we will aim to reproduce it in our holographic SQUID, and the result is \textcolor{black}{given in} Fig.\ref{Imax}. 

\subsection{Our holographic SQUID}
\label{SubChap:Review_of_hSQUID}

Now that we have reviewed the condensed matter physics side, we will turn to the points in our way to model the holographic SQUID, 
which are how to involve the magnetic flux and the flow of the supercurrent.   
Explicit descriptions of our model \textcolor{black}{are} presented after this section. 
\newline

\begin{figure}[h]
\begin{center}
\includegraphics[scale=0.71]{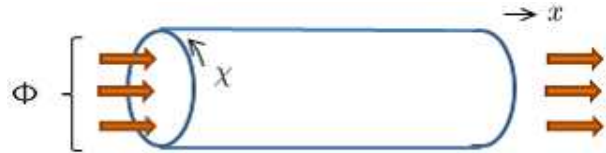}
\caption{
This figure illustrates the boundary space on which the dual field theory \textcolor{black}{lives}. 
\textcolor{black}{Here} our boundary space is given not as the cylinder including its interior, but as the surface of the cylinder.
We refer to the circled direction as $\chi$-direction.   
\textcolor{black}{Then} we put our holographic SQUID on $\chi$-direction in the form of the loop.
$\Phi$ represents the magnetic flux running in the interior of the cylinder, 
which itself is fictitious~(due to the absence of the interior space) 
but finally we can involve the effect of the magnetic flux $\Phi$ 
by the rewriting of the surface integral of the magnetic field to 
the contour integral of the gauge field~\cite{Montull:2012fy,Cai:2012vk}  
as mentioned in the body text.}
\label{bdspace}
\end{center}
\end{figure}

The background geometry in this paper will be the four-dimensional Anti-de Sitter Schwarzschild black brane geometry. 
\textcolor{black}{Then} the boundary space is given by 1+2 dimensional space. 
We compactify a direction of either of the two space directions into a $S^1$ circle, 
which means that the boundary space in this paper is given as the surface of the cylinder. 
We illustrate the boundary space in Fig.\ref{bdspace} on which the dual field theory lives. 
%
%
%
%

We realize our holographic SQUID on the $S^1$ circled space in the form that it winds around the $S^1$ circle.
The SQUID is composed of two Josephson junctions as can be seen in Fig.\ref{FigSQUID}.   
\textcolor{black}{Then} to model such double holographic Josephson junctions,  
we arrange the time component of the gauge field appropriately by exploiting its $\chi$ dependence as in eq.(\ref{mu}) as well as Ref.\cite{Horowitz:2011dz}, 
where the coefficients appearing in the expansion around the horizon are known to have the roles of the density~(charge) and the chemical potential. 
(For example, see Ref.\cite{Nakamura:2007nk}). 
In what follows, we refer the circled direction as $\chi$-direction.  
For more concrete description \textcolor{black}{of our model}, see Section.\ref{Chap:setup}.

\textcolor{black}{Here} we mention one of the issues in our holographic model, which is how to involve the magnetic flux. 
First, the space where the dual field theory lives does not include the interior space of the cylinder.  
Hence, the magnetic flux penetrating the circuit of the SQUID cannot exist in the dual field theory.  
However, temporarily considering the interior space of the cylinder in the space of the dual field theory, 
let us assume that the magnetic flux $\Phi$ given by the area integral $\displaystyle \Phi = \int d {\bf S} \cdot {\bf B}$ exists.  
\textcolor{black}{Then} considering the external gauge field $a_\chi(\chi)$ in the space of the dual field theory on the surface of the cylinder,
$\Phi$ can be rewritten to the line integral using the Stokes's theorem as $\displaystyle \Phi = \oint_\chi d\chi \, a_\chi(\chi)$.  
\textcolor{black}{Here} we write the coefficient of the $\chi$ component of the gauge field 
when it is expanded in the vicinity of the boundary in the bulk gravity as $\nu(\chi)$. 
\textcolor{black}{Then} we can link the magnetic flux $\Phi$ \textcolor{black}{that} 
we have spuriously assumed above to the external gauge field $\nu(\chi)$ actually existing in our model as 
\begin{eqnarray}\label{Phichi}
\Phi = \oint_\chi d\chi \, \nu(\chi).  
\end{eqnarray}
Hence, in the conclusion, despite \textcolor{black}{the fact} we cannot have the magnetic flux itself,  
we can fictitiously involve the effect of the magnetic flux. 
(This \textcolor{black}{way has} been performed in refs.\cite{Montull:2012fy,Cai:2012vk}.)
\newline

We have another issue, which is the effect of the branches \textcolor{black}{appearing} in Fig.\ref{FigSQUID}.  
First, since there is no branch in our holographic model of the SQUID, our holographic SQUID is just a loop \textcolor{black}{consisting} of two Josephson junctions.   
If there \textcolor{black}{are} the branches, the supercurrent inflows from the above, separates into two flows, and finally outflows to the below. 
On the other hand, if it were not for the branches, 
the circuit of the SQUID would become a simple loop and the supercurrent simply \textcolor{black}{circulates} in the circuit of the SQUID.

\begin{figure}[h]
\begin{center}
\includegraphics[scale=0.65]{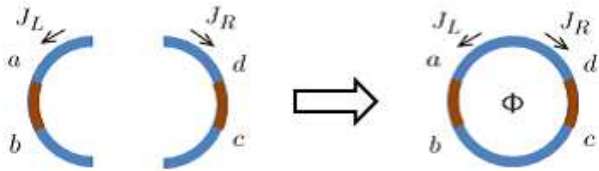}
\caption{
This figure illustrates that we simply consider the SQUID as the one except for the branches, and separate it into the left and the right parts.  
Then performing the analysis in each side \textcolor{black}{one at a time} separately, we finally consider the result of a SQUID by simply joining the results of these two analysis.   
}
\label{FigSQUID2}
\end{center}
\end{figure}

If we try to involve the effect of the branches, we have to consider the boundary condition such that all the fields in three sectors,  
the left, the right and the branch parts, \textcolor{black}{continuously connect to} each other at the joint parts between the circuit and the branches. 
However, such a boundary condition is very difficult to \textcolor{black}{treat}. 
\textcolor{black}{Hence in} this paper, considering that there is no \textcolor{black}{continuity condition} at the joint parts in the circuit of the SQUID in Fig.\ref{FigSQUID},   
the SQUID we consider is the one without the branches and separated into the left and the right parts as in the left figure of Fig.\ref{FigSQUID2}, 
where the meaning of the separation is mentioned in what follows.

First we mention the reason for \textcolor{black}{having} no \textcolor{black}{continuity condition}. 
Since the supercurrent suddenly splits \textcolor{black}{at the top aof the joint part and merges at the bottom joint part},  
the amount of the supercurrent varies rapidly at the joint parts, which is mostly discontinuous.  
Therefore, we can consider that the fields at the joint parts are \textcolor{black}{discontinuous}. 
\textcolor{black}{Then} we mention about separating into the left and the right parts in Fig.\ref{FigSQUID2}.  
Since there is no \textcolor{black}{continuity condition} as mentioned above, we can consider that there is no interference between the two Josephson junctions.  
Therefore, we perform the analysis not for the two Josephson junctions interfering \textcolor{black}{with} each other but for each side of the Josephson junction separately.

\textcolor{black}{After this}, we \textcolor{black}{reference} how to obtain the result of the SQUID from \textcolor{black}{two such} independent Josephson junctions.
First, we assume that we have obtained the results for one Josephson junction with no \textcolor{black}{interaction} with the other Josephson junction. 
Actually, the result for one Josephson junction is presented in Fig.\ref{sine_relation}, 
and the numerical data on which Fig.\ref{sine_relation} is based is presented in Table.\ref{nnn} of \ref{app:2}, 
\textcolor{black}{Then} we choose the value of the supercurrents from Table.\ref{nnn} and simply treat these as the amount of the supercurrent flowing in the each side. 
Accompanying this, the phase differences in each Josephson junction are chosen. 
\textcolor{black}{Then} from the value of the supercurrents and the phase differences we have set here, 
according to eq.(\ref{twopin2}), the magnetic flux $\Phi$ can be fixed.   
\textcolor{black}{Then} now that we have the information of the supercurrents in each side and the magnetic flux $\Phi$, 
we can obtain the values of $J_{\rm max}(\Phi)$ in eq.(\ref{totalcurrent2}), 
and finally obtain the results of the SQUID, which is Fig.\ref{FigSQUID2}.

\section{Holographic setup to model SQUID}
\label{Chap:setup}

In this paper, \textcolor{black}{we consider} the following action
\begin{eqnarray}\label{action}
S=\int d^4 x \; \sqrt{-g}\left[R+\frac{6}{L^2}-\frac{1}{4}F_{\mu\nu}F^{\mu\nu}-|D\psi|^2-m^2|\psi|^2\right], 
\end{eqnarray}
where $\mu,\nu=1,\cdots,4$, $A_\mu$ \textcolor{black}{is U(1)} gauge field and $F_{\mu\nu}=\partial_\mu A_\nu-\partial_\nu A_\mu$.  
Further, $D_\mu\equiv \nabla-iqA_\mu$ and $m^2=-2$. 
\textcolor{black}{We take} the probe approximation in our analysis,  
which can be obtained by rescaling $\psi=\tilde\psi/q$, $A=\tilde A/q$ and taking $q\rightarrow\infty$ 
with \textcolor{black}{fixing $\tilde\psi$ and $\tilde A$}.

One of \textcolor{black}{the} solutions \textcolor{black}{in model} (\ref{action}) is the four-dimensional Anti-de Sitter Schwarzschild black brane geometry, 
\begin{eqnarray}\label{metric}\textcolor{black}{
ds^2 = \frac{1}{z^2}\left(-f(z) dt^2 + \frac{1}{f(z)}dz^2\right) + r^2(dx^2+d\chi^2),}
\end{eqnarray}
where $f(r) \textcolor{black}{=} 1-z^3/z_0^3$ \textcolor{black}{and $z_0$ mean the location of the horizon}.
$\chi$-direction is $S^1$ compactified with the periodicity $-\pi R \leq \chi\leq \pi R$ with $\pi R=7$ 
~(For convenience in our actual analysis, we \textcolor{black}{have taken this a number}).   
\textcolor{black}{Here} the Hawking temperature is given as \textcolor{black}{$T=3/(4\pi L^2z_0)$}.  
We \textcolor{black}{fix the AdS radius $L$ and $r_0$ as $L=1$ and $z_0=1$, respectively}. 
As a result, the temperature \textcolor{black}{in this paper is given as} $T=3/(4\pi)$.

\textcolor{black}{Here} let us give a comment on another geometry. The soliton geometry~\cite{Horowitz:1998ha} is also one of solutions in the model (\ref{action}). 
However, in the case of the soliton geometry, due to the vacuum current and the difference in the fluxoid number~\cite{Montull:2012fy},  
the analysis will be  \textcolor{black}{very} involved. 
\textcolor{black}{Therefore in} this paper, we \textcolor{black}{fix} the background geometry to the black brane geometry, 
and the boundary space is given as \textcolor{black}{the} surface of a cylinder.   
\textcolor{black}{Here} it is \textcolor{black}{known~\cite{Nishioka:2009zj}} that 
the black brane geometry (\ref{metric}) is energetically favorable \textcolor{black}{towards} the soliton geometry 
provided that $T>1/(2\pi R)$  \textcolor{black}{with $R$ taken as $R=7/\pi$ in this paper.}

\textcolor{black}{As an ansatz, we think that all the fields can be written as follows:}
\textcolor{black}{
\begin{eqnarray}
\tilde\psi(z,\chi) &=& \overline{\psi}(z,\chi)e^{i\varphi(z,\chi)},\\
\tilde A(z,\chi) &=& A_t(z,\chi) dt + A_z(z,\chi) dz + A_\chi(z,\chi) d\chi,
\end{eqnarray}
}
where $|\psi|$, $\varphi$, $A_t$, \textcolor{black}{$A_z$}, and $A_\chi$ are realistic function\textcolor{black}{s} of $r$ and $\chi$ 
and \textcolor{black}{the} periodic for $\chi$-direction as \textcolor{black}{$\tilde\psi(z,\chi+14) = \tilde\psi(z,\chi)$} and 
\textcolor{black}{$\tilde A(z,\chi+14) = \tilde A(z,\chi)$}, \textcolor{black}{but} not continuous at $\chi=0, \pm 7$ for the reason 
mentioned in subsection.\ref{SubChap:Review_of_hSQUID}, where the branches as in Fig.\ref{FigSQUID} attach at $\chi=0, \pm 7$.   
In the following, we \textcolor{black}{perform the analysis} with the gauge-invariant quantity $M_\mu \equiv A_\mu-\partial_\mu \varphi$ 
as well as Ref.~\cite{Horowitz:2011dz}.

%
%

\subsection{\textcolor{black}{Equations of motion and solutions around the horizon and the boundary}}
\label{SubChap:sol_and_bc}
 
We can obtain the equations of motion as 
\textcolor{black}{
\begin{subequations}
\begin{align} 
\label{eom1a}
2 \, z^{1-\sqrt{4 {m^2}+9}} \, \overline{\psi}\,{}^2 \, M_t - \partial_\chi^2 M_t - f\,\partial_z^2 M_t&= ~0,\\
\label{eom2a}
2 \, z^{1-\sqrt{4 {m^2}+9}} \, \overline{\psi}\,{}^2 \, M_z  - \partial_\chi^2 M_z - \partial_z \partial_\chi M_\chi&= ~0,\\
\label{eom3a}
2 \, z^{1-\sqrt{4 {m^2}+9}}\, \overline{\psi}\,{}^2 \, M_\chi + 3 \, z^2 \, \partial_\chi M_z - f\,\partial_z \partial_\chi M_z + {3 \, z^2 \partial_z M_\chi} - f\,\partial_z^2 M_\chi&= ~0, \\
\label{eom4a}
\left\{ \left(\sqrt{4 {m^2}+9}-4\right) z^3 - \sqrt{4 {m^2}+9}+1 \right\} \partial_z \overline{\psi} +\frac{1}{2}\left(-2 {m^2}+3 \sqrt{4{m^2}+9}-9\right)\,z^2 \,\overline{\psi} \nonumber\\
- z f\,\overline{\psi}\,M_z^2 + \frac{z}{f}\, \overline{\psi}\,M_t^2 - z\, \overline{\psi} \, M_\chi^2 + z \, \partial_\chi^2 \, \overline{\psi} + z \,f \, \partial_z^2 \overline{\psi}&= ~0,
\\
\label{eom5a}
\left\{\left(\sqrt{4 {m^2}+9}-4\right) z^3 - \sqrt{4{m^2}+9} + 1 \right\} \overline{\psi} \, M_z \nonumber\\ 
+ z \left(1-z^3\right)\overline{\psi} \, \partial_z M_z + 2\,z \, f \, M_z \, \partial_z \overline{\psi} - z \, \overline{\psi} \, \partial_\chi M_\chi - 2\,z \,M_\chi \, \partial_\chi \overline{\psi}&= ~0.
\end{align}
\end{subequations}
}
\textcolor{black}{Here} the \textcolor{black}{above equations} are equations of motion \textcolor{black}{regarding the} fields associated with $|\psi|$, 
and the equation of motion with regard to $A_x$ decouples. 
Further, \textcolor{black}{it turned} out that there is a relation~:~
\textcolor{black}{
$ \displaystyle
- z^2 \partial_z \left( z^2 \, f \cdot {\rm eq}.(\ref{eom2a}) \right) 
+ z^4 \, \partial_\chi {\rm eq}.(\ref{eom3a}) 
+ 2z^{-1-\sqrt{9+4m^2}}\,\overline{\psi} \cdot {\rm eq}.(\ref{eom5a})  =0
$}. 
Hence, both the number of independent equation\textcolor{black}{s} in the above and the number of variable\textcolor{black}{s} appearing in the above equations are four.

We can see that the equations of motion (\ref{eom1a})-(\ref{eom5a}) are invariant under the following rescaling, 
\textcolor{black}{\begin{eqnarray}\label{scaling}
&&
(t,~\chi,~\textcolor{black}{x},~r) \to (t/a,~\textcolor{black}{\chi}/a,~\textcolor{black}{x}/a,~a \, r),\\
&&(M_t,~M_\chi,~M_{\textcolor{black}{x}},~M_r) \to (M_t/a,~M_{\textcolor{black}{\chi}}/a,~M_{\textcolor{black}{x}}/a,~a \, M_r), 
\end{eqnarray}}
where $a$ is a rescaling parameter. 
In our actual analysis, we fix this scale invariance \textcolor{black}{so} that \textcolor{black}{$z_0$ becomes $1$}.

\textcolor{black}{Then} the expansions of the solutions near the boundary turn out to be given as 
\begin{subequations}
\begin{align}
\label{asyz0a}  
\overline{\psi}(z,\chi)
&=
\overline{\psi}^{(1)}(\chi)
+
\overline{\psi}^{(2)}(\chi) \, z^{\sqrt{9+4m^2}}
+
\mathcal{O}(z^{2\sqrt{9+4m^2}}),
\\
\label{asyz0b} 
M_t(z,\chi)
&=
\mu(\chi)
-
\rho(\chi) \, z
+
\mathcal{O}\left( z^2 \right),
\\
\label{asyz0c} 
\textcolor{black}{M_z(z,\chi)}
&=
\textcolor{black}{M_z^{(1)}(\chi) \, z+\mathcal{O}(z^2),}
\\
\label{asyz0d} 
M_\chi(z,\chi)leads
&=
\nu(\chi)
+
J(\chi) \, z 
+
\mathcal{O}\left( z^2 \right).
\end{align}
\end{subequations}
\textcolor{black}{Here} in the dual field theory, $\mu(x)$ and $\rho(x)$ have the roles as the chemical potential and the density for the Cooper pair, respectively. 
On the other hand, $\nu(x)$ and $J(x)$ have the roles as the gauge field associated with the magnetic flux as in eq.(\ref{Phichi}) and the supercurrent of the Cooper pair, \textcolor{black}{respectively}. 
We take up $\mu(x)$ in more detail later again.
Next, with \textcolor{black}{regards} to $\overline{\psi}^{(1)}(\chi)$ and $\overline{\psi}^{(2)}(\chi)$, 
the existence of $\overline{\psi}^{(1)}(\chi)$ leads \textcolor{black}{to} the term $\overline{\psi}^{(1)}(\chi) \, \overline{\psi}^{(2)}(\chi)$ in the field theory side of the GKP-W relation~\cite{Gubser:1998bc,Witten:1998qj}, 
and the existence of that term breaks the U(1) global symmetry in the dual field theory.  
For this reason, we have taken $\overline{\psi}^{(1)}(\chi)$ to zero. 
\textcolor{black}{Then} $\overline{\psi}^{(2)}(\chi)$ will have the role of the wave function \textcolor{black}{for} the Cooper pair in the boundary theory.

\begin{figure}[h]
\begin{center}
\includegraphics[scale=0.50]{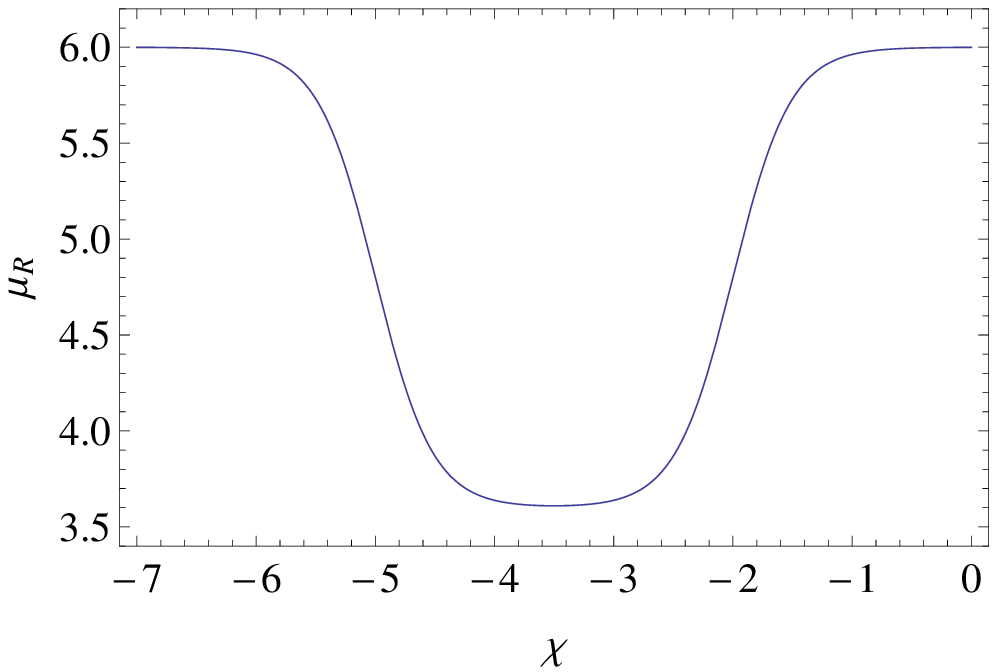}
\includegraphics[scale=0.50]{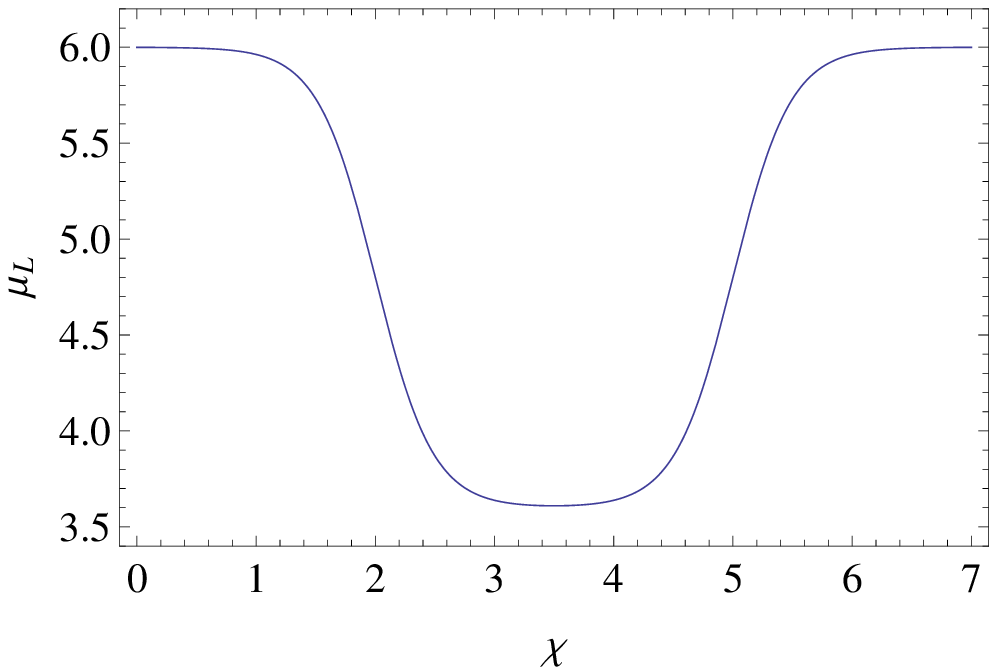}
\caption{
An example 
of the profile 
of the chemical potential 
given in eq.(\ref{mu}) with the parameters:~
$\mu_H = 6.0$,~
$\kappa = 6.5$,~ 
$\delta = 3.5$,~
$\epsilon = 1.5$ 
and 
$\lambda=1.2$.  
\textcolor{black}{Here} 
$\chi$-direction is $S^1$ compactified with the period $14$ and \textcolor{black}{$m^2=-2$}.  
Although 
we show two figures here, 
these are indeed connected 
in $S^1$ circled space, 
where we assume 
the branch parts 
as in Fig.\ref{FigSQUID} 
locate at $\chi=0$ and $\pm 7$, 
from which the supercurrents flow in and flow out. 
Why we show a \textcolor{black}{single} profile 
as two profiles separately is that 
we perform the analysis 
for each one independently 
for the reason mentioned 
in section \ref{SubChap:Review_of_hSQUID}. 
\textcolor{black}{Here}  
notice that,  
taking into \textcolor{black}{account the} fact that 
we measure the phases 
in \textcolor{black}{an} anticlockwise direction in Fig.\ref{FigSQUID}, 
the right and the left figures correspond 
to the chemical potentials 
in the left and the right parts 
in the SQUID. 
}
\label{shape_mu}
\end{center}
\end{figure}

\textcolor{black}{Then} let us turn to $\mu(\chi)$ in more detail. 
It is generally known in the GKP-W relation~\cite{Gubser:1998bc,Witten:1998qj} that $\mu(\chi)$ has a role of the chemical potential for the Cooper pair, 
and we \textcolor{black}{assign  the following profile to $\mu(\chi)$~:}
\begin{align}\label{mu}
\mu(\chi) = \mu_L(\chi) + \mu_R(\chi)
\end{align}
with
\begin{align}\label{mu1}
\mu_L(\chi)= \mu_H &-
\lambda \left[ \tanh \left\{ \frac{\kappa (\chi-\delta+\epsilon)}{\pi} \right\} - \tanh \left\{ \frac{\kappa (\chi-\delta-\epsilon)}{\pi} \right\} \right],\\
\mu_R(\chi) = \mu_H &- \lambda \left[ \tanh \left\{\frac{\kappa (\chi+\delta+\epsilon)}{\pi}\right\} - \tanh \left\{\frac{\kappa (\chi-\delta-\epsilon)}{\pi}\right\}\right].
\end{align}
\textcolor{black}{Here} \textcolor{black}{$\kappa$, $\delta$, $\epsilon$, $\lambda$ and $\mu_H$} are parameters to fix the profile of $\mu_L(\chi)$ and $\mu_R(\chi)$, 
where $\kappa$ has the role of roundness, $\delta$ has the role of position, $\epsilon$ has the role of \textcolor{black}{width} 
$\lambda$ has the role of depth and $\mu_H$ has the role of height.  
The profile of $\mu(\chi)$ in this paper is shown in Fig.\ref{shape_mu}, 
in which we can see that there \textcolor{black}{is} a height difference in the profile.  
In the following several paragraphs, we mention why we take the profile \textcolor{black}{in} such a form.

We have set temperature to $T=3/(4 \pi)$ using the rescaling given in eq.(\ref{scaling}).  
\textcolor{black}{Then} the effect of \textcolor{black}{the} temperature comes from either one of the ratio\textcolor{black}{s} $T/\mu$ or $T/T_{\rm c}$, 
where $T_{\rm c}$ means the critical temperature for the superconductor/normal metal transition in our model.
we can take in the effect of temperature not from the ratio of $T/\mu$ but from  the ratio $T/T_{\rm c}$. 
\textcolor{black}{Here} let us show the critical temperatures in our paper.

First, there are the higher and the lower sections in the profile of our chemical potential (\ref{mu})~(and Fig.\ref{shape_mu}).
\textcolor{black}{Next}, \textcolor{black}{by} denoting the values of each section as $\mu_{\rm H}$ and $\mu_{\rm L}$, 
the critical temperatures for each section are known to be given as
$\displaystyle T_{\rm cH} =  c \, \mu_{\rm H}$ and $\displaystyle T_{\rm cL} =  c \, \mu_{\rm L}$ with 
$\displaystyle c = 0.0588$~\cite{Horowitz:2011dz}, 
where $\displaystyle T_{\rm cH}$ and $\displaystyle T_{\rm cL}$ mean the critical temperatures in the higher and the lower sections, respectively.

\textcolor{black}{Then} when \textcolor{black}{the} temperature is higher or lower than the critical temperature, 
the phase of that section is the normal metal or the superconductor, respectively. 
Hence, in order to model a Josephson junction holographically, since temperature $T$ has been fixed to $T=3/(4 \pi)$ as mentioned above,
we should set the critical temperatures $T_{\rm cL}$ and $T_{\rm cH}$ such that 
$\displaystyle 
T_{\rm cL} < T < T_{\rm cH}$ for fixed $T$, which means that we should assign some different values to $\mu_{\rm H}$ and $\mu_{\rm L}$ \textcolor{black}{so} that 
these satisfy the relation     
$\displaystyle
\frac{\mu_L}{\mu_H} < \frac{T}{c\,\mu_H} < 1,
$
and this is why we have made the height difference in the profile of the chemical potential, 
where we have used the relation mentioned above:~$\displaystyle T_{\rm cH} =  c \, \mu_{\rm H}$ and $\displaystyle T_{\rm cL} =  c \, \mu_{\rm L}$.

\textcolor{black}{The} actual calculations in this paper \textcolor{black}{are} always set $\mu(\chi)$ as \textcolor{black}{shown} in Fig.\ref{shape_mu}, 
where we assume in the figure that the branch parts \textcolor{black}{presented} in Fig.\ref{FigSQUID} locate at $\chi=0$ and $\pm 7$. 
\textcolor{black}{Here} let us notice that in Fig.\ref{shape_mu} taking into account \textcolor{black}{the} fact that 
we measure the phases  in \textcolor{black}{an} anticlockwise direction in Fig.\ref{FigSQUID}, the right and the left 
figures correspond to the chemical potentials in the left and the right parts in the SQUID. 
By setting so, we perform our analysis for each side \textcolor{black}{one at a time} separately 
\textcolor{black}{giving} the various values of supercurrents flowing in the left and the right sides as the initial values. 
It means that, there being \textcolor{black}{a} stage where the supercurrent \textcolor{black}{flows} into the left and the right sides \textcolor{black}{is} 
determined \textcolor{black}{by} the configuration of the Josephson junctions in the left and the right sides and the amount of the supercurrent flowing into the SQUID,  
such a stage  is skipped in our analysis.  
We have mentioned the validity for this skip in section \ref{SubChap:Review_of_hSQUID}.

\textcolor{black}{By} simply joining the results of each Josephson junction, we read out the results as the results of a SQUID.
In \textcolor{black}{the} \textcolor{black}{following}, 
let us mention the condition that the profile of the chemical potential 
should satisfy to model a Josephson junction. 

\section{The analyses and the results}
\label{analysis}

\begin{figure}[h]
\includegraphics[scale=0.2710]{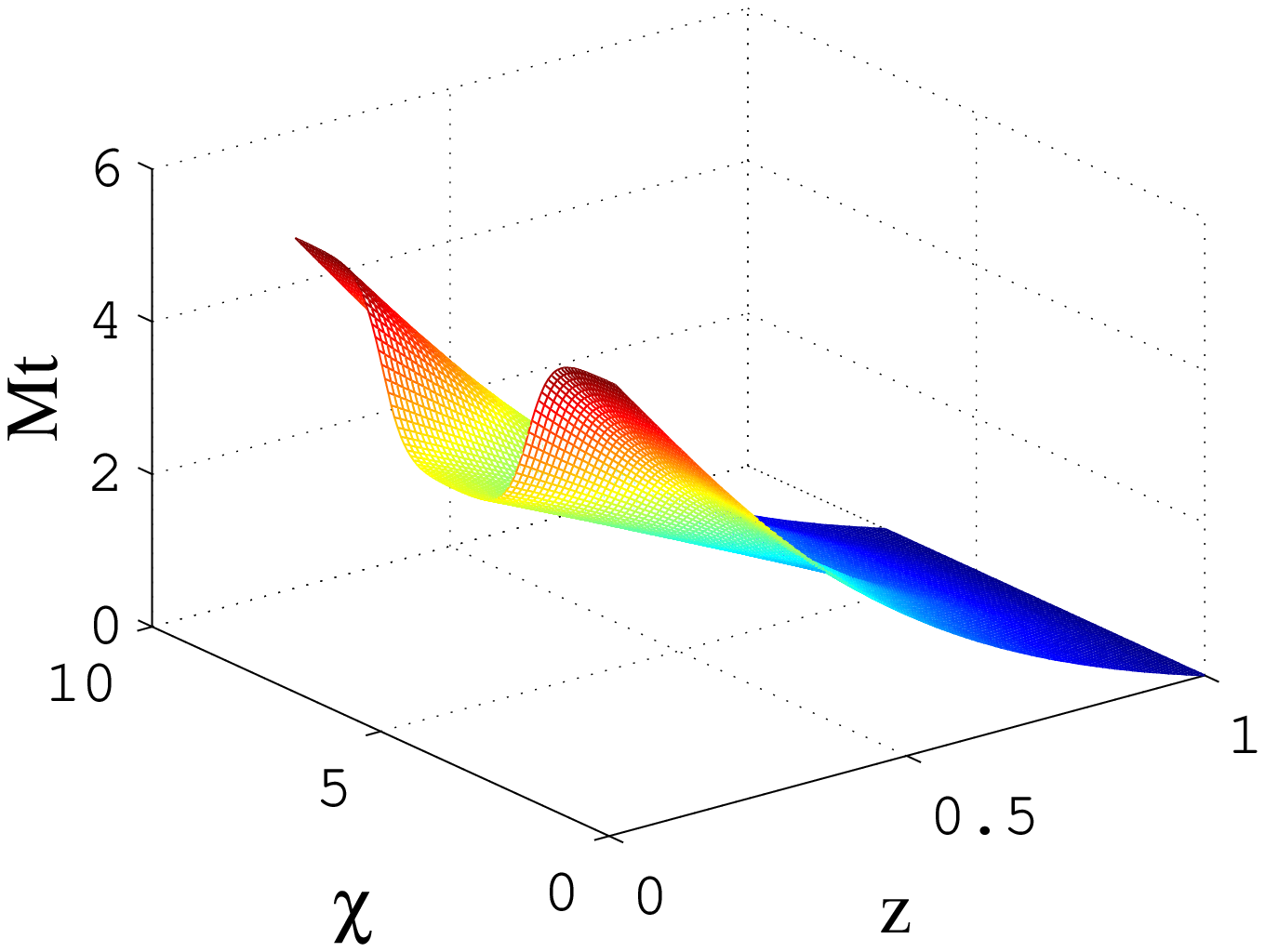}     
\includegraphics[scale=0.2710]{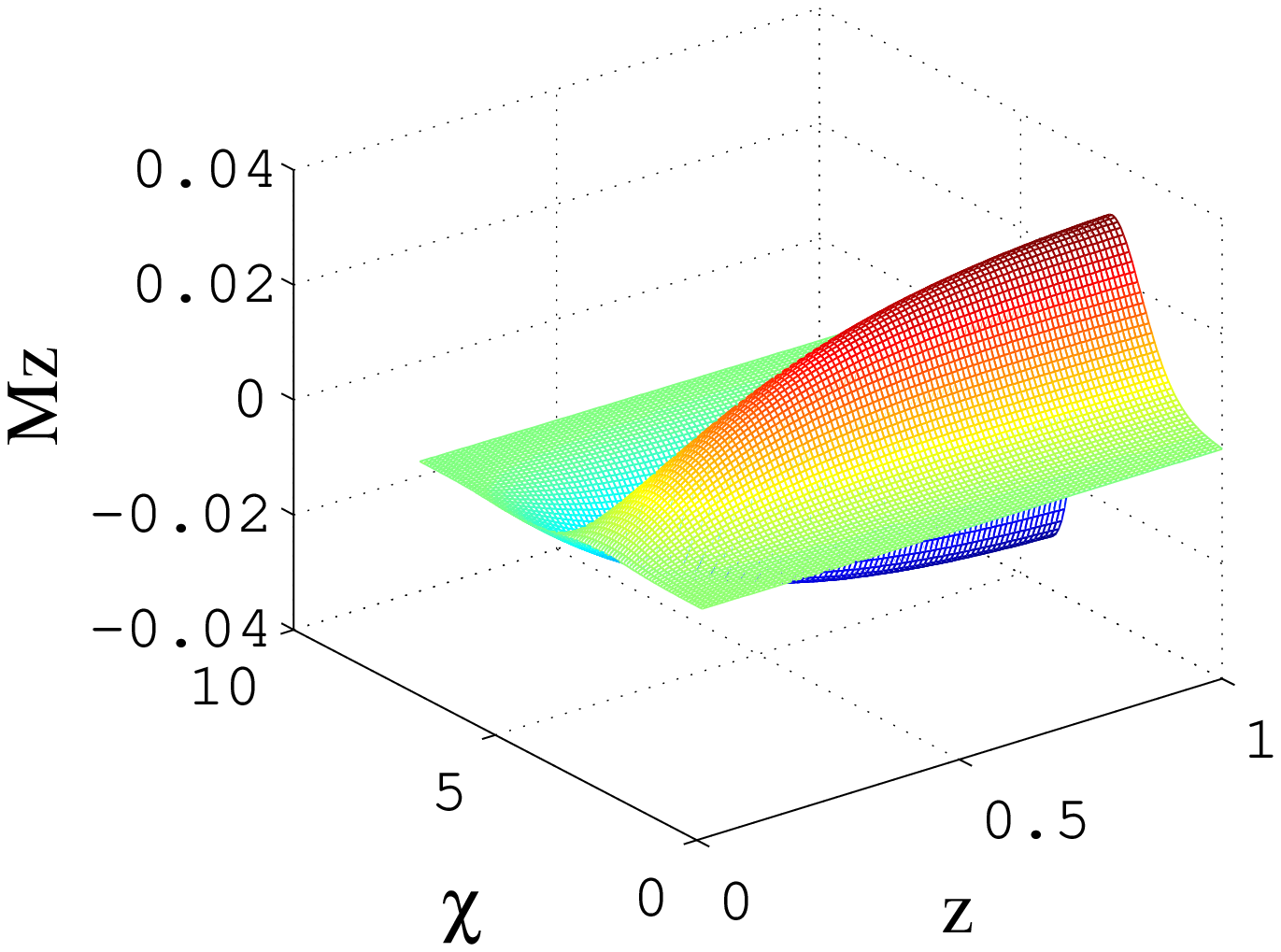}     
\includegraphics[scale=0.2710]{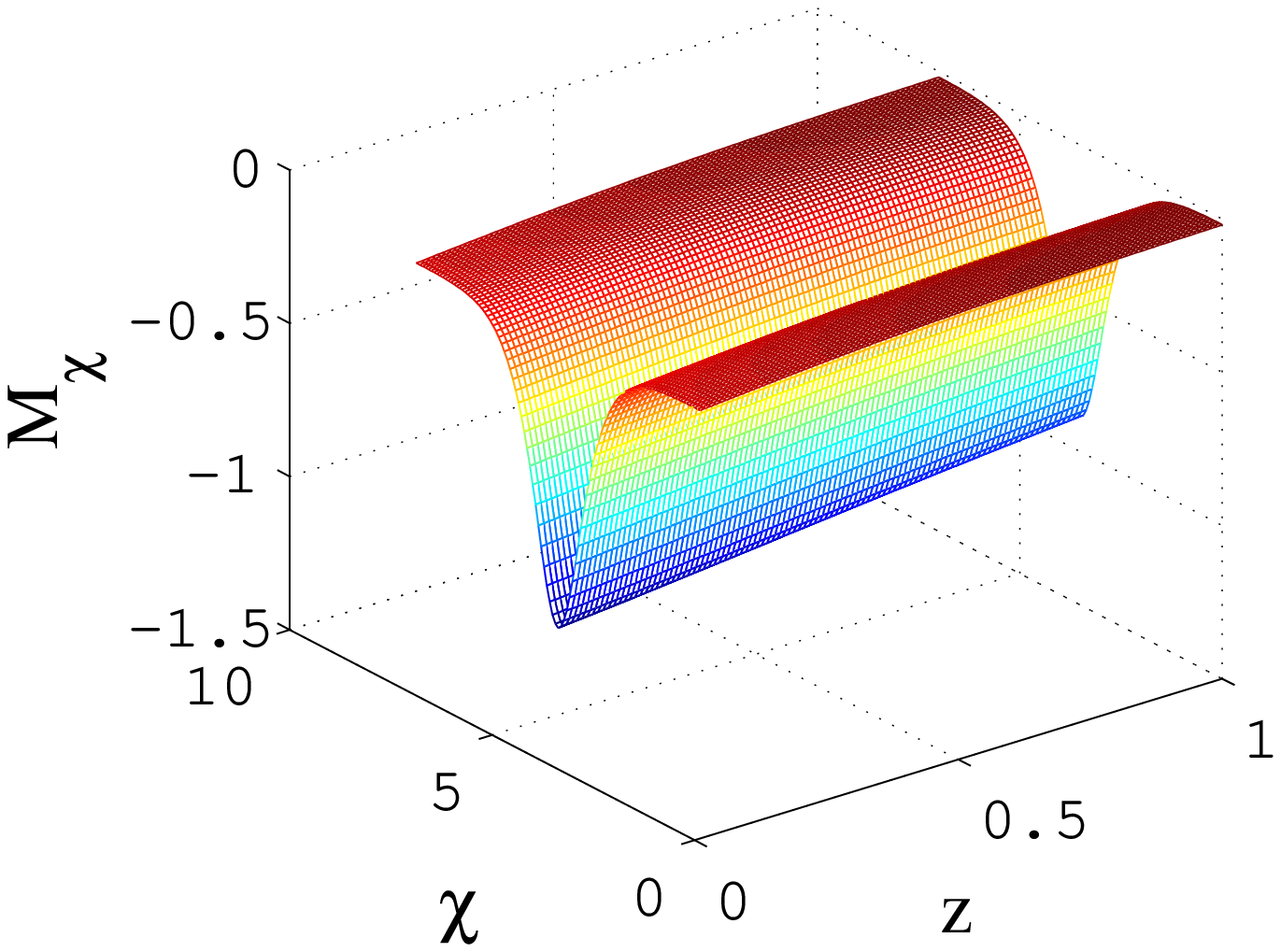}  
\includegraphics[scale=0.2710]{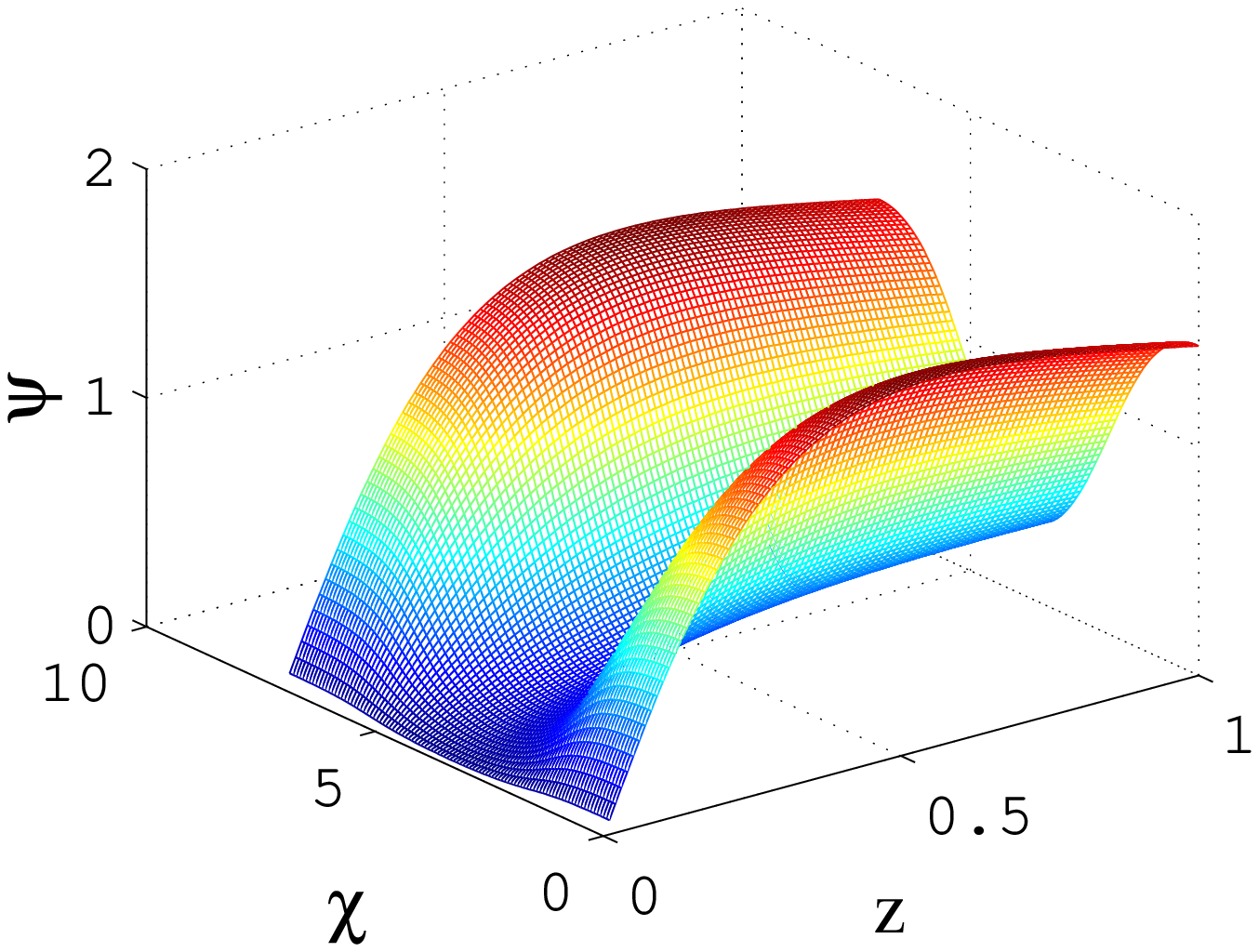}    
 \caption{
These figures represent examples 
for the solutions of 
$M_t$, $M_r$ $M_\chi$ and $|\psi|$ 
obtained in the parameters as follows: 
$J/T_{\rm cH}^2=1.21316$, 
the chemical potential (\ref{mu}) taken in this calculation is same with the one in the right figure of Fig.\ref{shape_mu}\textcolor{black}{,}  
the Chebyshev Grid is taken as $(n_z,n_\chi)=(23,45)$, 
where $n_z$ and $n_\chi$ mean number or the grid in $z$- and $\chi$-directions.  
\textcolor{black}{Here} these calculations are performed in the half of the whole $\chi$ space 
since our calculations are performed for the left and the right parts in the circuit of the SQUID \textcolor{black}{one at a time} separately.     
We show how the sections of the solutions of $|\psi|$ and $M_\chi$ at the boundary, $z=0$, 
in Fig.\ref{fields2}, which \textcolor{black}{means} the condensation of the cooper pair and $\chi$ component 
of the gauge field in the dual field theory,  respectively.   
}
\label{fields}   
\end{figure}

Taking the chemical potential $\mu$ 
for each side  separately\textcolor{black}{,} as in Fig.\ref{shape_mu}, 
we numerically solve 
the equations of motion (\ref{eom1a})-(\ref{eom5a}) twice 
with various $J$ 
as the inputs of the numerical calculation.   
Namely, 
our analysis is the one 
which performs 
two calculations for a Josephson junction. 
In our \textcolor{black}{solution}, 
we impose the boundary conditions $M_t|_{z=z_0}=0$ 
and that 
\textcolor{black}{$M_z$} is \textcolor{black}{an} odd function and $M_t$, $M_\chi$ and $|\Psi|$ are even functions 
for $\chi$-direction,  
where $\chi$-direction is \textcolor{black}{half the} space of the whole $\chi$ space, 
since 
we perform the calculation 
for each side of the chemical potential 
in each side of the circuit 
of the SQUID separately. 
To this purpose,  
we use the spectral method 
on the Chebyshev Grid~\cite{trefethn}.  
We show examples of the solutions 
we have obtained in Fig.\ref{fields}.  
\begin{figure}[h]
\begin{center}
\includegraphics[scale=0.2710]{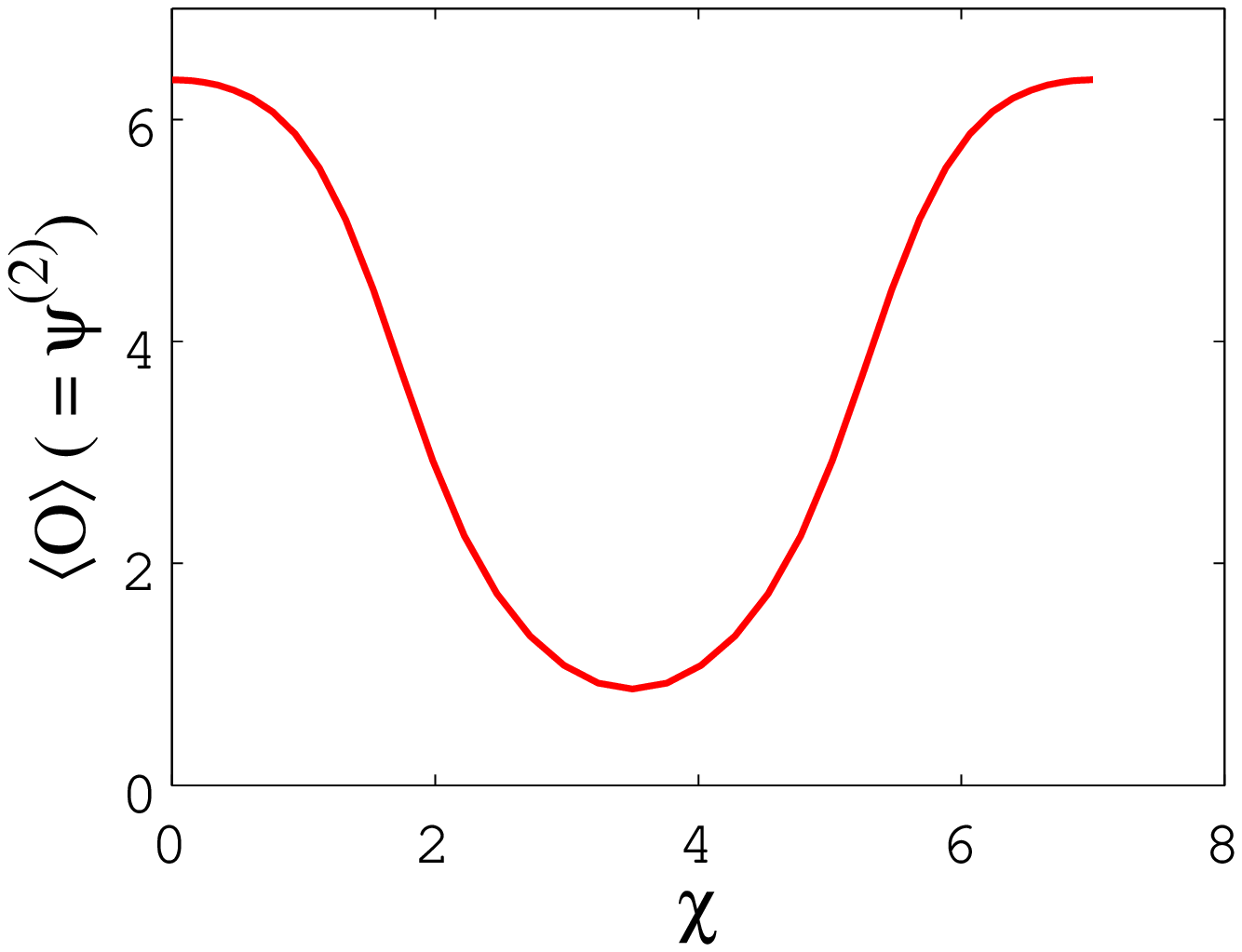}
\includegraphics[scale=0.2710]{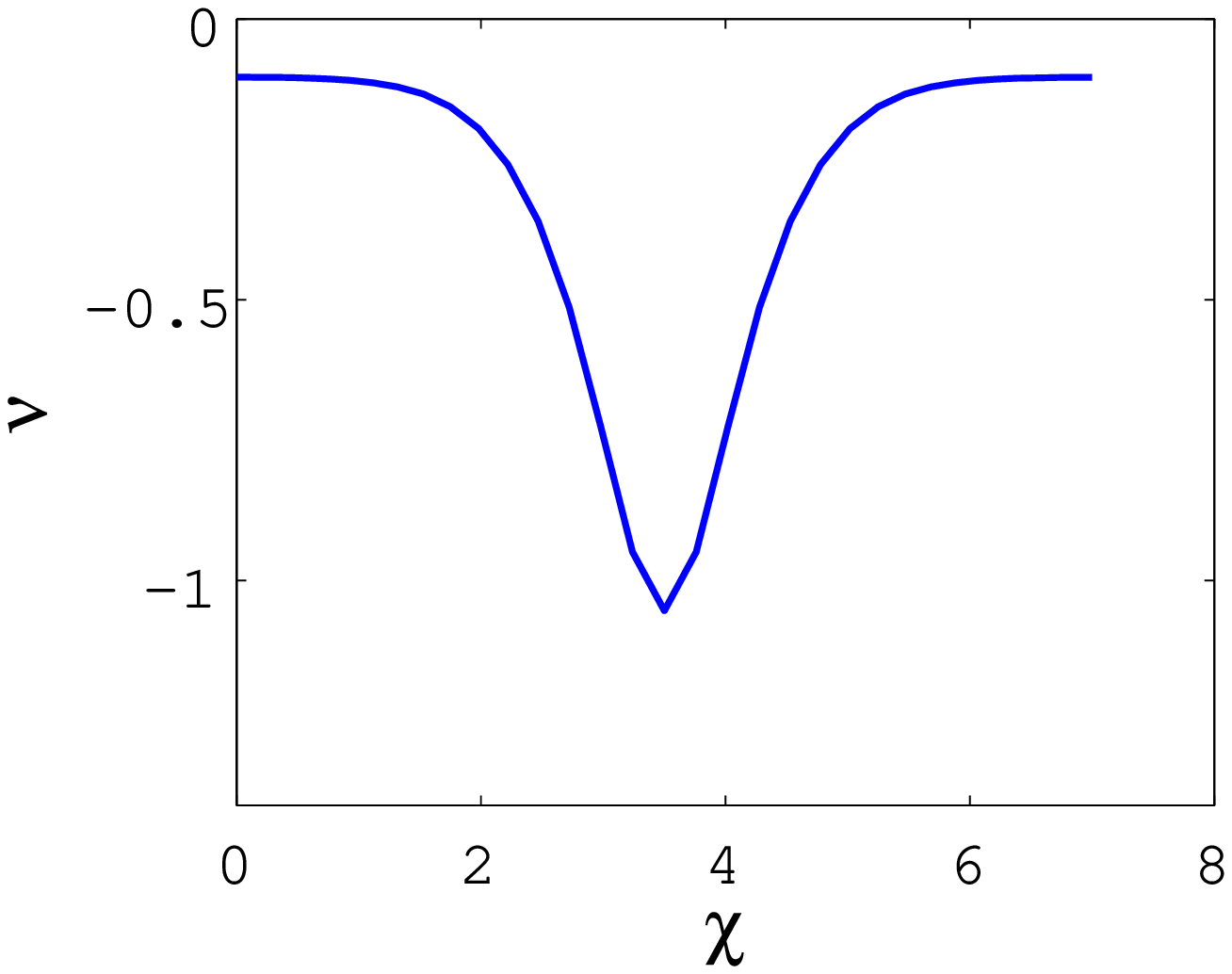}
 \caption{
These two figures represent the sections of the solutions of $\overline{\psi}^{(2)}$ and $M_\chi$ at the boundary, $z=0$, 
which mean the condensation of the cooper pair $\langle {\cal O }\rangle$ and $\chi$ component of the gauge field in the dual field theory, respectively.   
}
\label{fields2}   
\end{center}
\end{figure}

\begin{figure}[h]
\begin{center}
\includegraphics[scale=0.750]{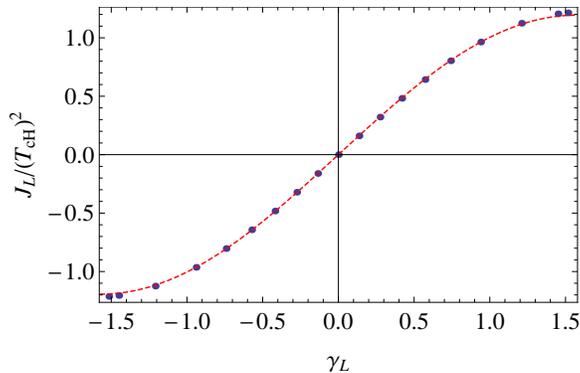}
 \caption{
The blue points \textcolor{black}{represent} our numerical result 
obtained from solving the equations of motion (\ref{eom1a})-(\ref{eom5a}) 
with the chemical potential $\mu$ 
given in the right figure 
\textcolor{black}{of} Fig.\ref{shape_mu} 
and various $J_L$ 
as the inputs of the numerical calculations.  
\textcolor{black}{Here} $x$- and $y$-\textcolor{black}{axis show} the phase difference $\gamma_L$ given in eq.(\ref{phasedifference}) 
and the supercurrent $J_L$ normalized by $T_{\rm cH}^2$, where $T_{\rm cH} \equiv c~\mu_H$ with $c \equiv 0.0588$~\cite{Horowitz:2011dz}. 
The dashed line is the \textcolor{black}{guide line to show} that these results are on a sine curve, which is $J_L/(T_{\rm cH})^2=1.1935 \sin (\gamma_L)$.
The result on a sine curve like this figure is one of the specific \textcolor{black}{behaviors} 
in a Josephson junction as mentioned in Section.\ref{SubChap:Review_of_SQUID}. 
We can see that the section of $x$-axis is from about $-\pi/2$ to $\pi/2$.  
}
\label{sine_relation}   
\end{center}
\end{figure}

Finally, we can obtain the numerical results shown in Fig.\ref{sine_relation}, where we list the numerical results of the calculations in Table.\ref{nnn}.  
The dashed line in the figure is the guide for the eye to show that our results are on a sine curve, 
which is $J_L/(T_{\rm cH})^2=1.1935 \sin (\gamma_L)$.
The result of this sine curve is one of the specific behavior in a Josephson junction as mentioned in Section.\ref{SubChap:Review_of_SQUID}.

\textcolor{black}{Here} we \textcolor{black}{indicate} how to measure the phase difference in this paper. 
\textcolor{black}{According to Ref.\cite{Horowitz:2011dz}, we define} the following phase difference 
for the left and the right sides respectively as
\begin{eqnarray}\label{phasedifference}
\gamma_L \equiv \int_{0}^7 d\chi \left( \nu(\chi)-\nu(0) \right) 
\quad {\rm and} \quad
\gamma_R \equiv \int_{-7} ^{0}
d\chi \left( \nu(\chi)-\nu(-7) \right). 
\end{eqnarray} 
\textcolor{black}{Here} we have taken into account of the fact that we measure the phases in \textcolor{black}{an} anticlockwise direction in the circuit of the SQUID in Fig.\ref{FigSQUID}. 
\newline

Having obtained the result in one Josephson junction, let us turn to the holographic SQUID.    
The holographic SQUID we consider is composed of two Josephson junctions made \textcolor{black}{from} the chemical potential given in Fig.\ref{shape_mu}.  
To begin with, let us use $J_L$ and $J_R$ to denote the supercurents 
flowing in the left and the right sides in the circuit of the SQUID in Fig.\ref{FigSQUID}, respectively.   
\textcolor{black}{Then} considering the circuit of the SQUID separately as in Fig.\ref{FigSQUID2}, 
we set various values of $J_L$ as in Table.\ref{nnn} with a fixed $J_R=-0.482052$. 
\textcolor{black}{Here} this $J_R$\textcolor{black}{is} flowing from the top to the bottom in Fig.\ref{FigSQUID}, 
since we define the $J_{\rm total}$ as in eq.(\ref{totalcurrent1}).   
Further, such a setting corresponds to the situation \textcolor{black}{
where the supercurrent flowing in the left side varies and the supercurrent in the right side flowing constantly.}

We have described the validity for giving the values of each supercurrent flowing in the left and the right sides by hand in subsection.\ref{SubChap:Review_of_hSQUID}.   
Since the values of the supercurrents are set, the phase differences are determined.  
\textcolor{black}{Then} the magnetic flux $\Phi$ is determined from Table.\ref{nnn} according to eq.(\ref{twopin2}).
As a result, from the information of the values of the supercurrents and the magnetic flux, 
we can read out the relation between the total current $J_{\rm total}$ given in eq.(\ref{totalcurrent1}) 
and the magnetic flux $\Phi$ induced by the supercurrents flowing in the circuit.

\begin{figure}[h]
\begin{center}
\includegraphics[scale=0.800]{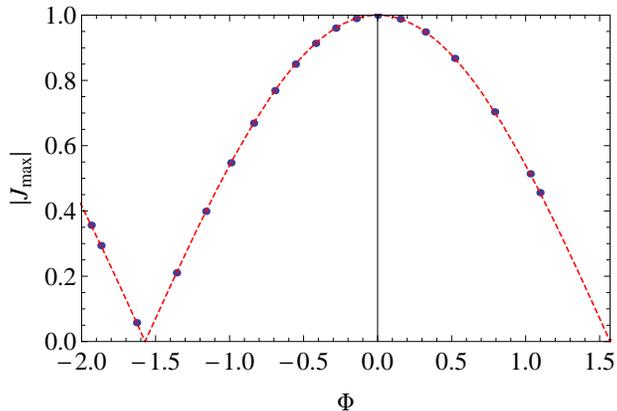}
 \caption{
The blue points \textcolor{black}{represent} our numerical result 
for the relation between the magnetic flux $\Phi$~(x-axis) and the absolute value of the maximum amplitude of the supercurrent $|J_{\rm max}|$ 
flowing into the circuit~(y-axis). As can been seen from eq.(\ref{totalcurrent2}), 
$|J_{\rm max}|$ is to be given as a cosine curve, and the dashed line is the \textcolor{black}{guideline} 
to show that our results are on an absolute cosine curve.
The result \textcolor{black}{of} an absolute cosine curve like this figure is one of the specific behaviors in the SQUID.  
In the calculation \textcolor{black}{is} the current flowing from the top to the bottom in Fig.\ref{FigSQUID}, 
since we define the $J_{\rm total}$ as in eq.(\ref{totalcurrent1})), 
which means that \textcolor{black}{the supercurrent flowing in the left side varies and the supercurrent in the right side flowing constantly.} 
\textcolor{black}{Here} we can see the disparity in this figure, which is \textcolor{black}{between the left and the right in the positions of the blue points}.  
The reason for that disparity is simply that \textcolor{black}{the} actual numerical data obtained in each Josephson \textcolor{black}{junction,} 
as in Table.\ref{nnn}\textcolor{black}{, is} from about $-\pi/2$ to $\pi/2$, 
and the amount of the supercurrent in one side is fixed to a finite value. 
}
\label{Imax}   
\end{center}
\end{figure}

As a result, we can obtain the maximum amplitude of the supercurrent flowing into the circuit $J_{\rm max}$\textcolor{black}{,} 
given in eq.(\ref{totalcurrent2})\textcolor{black}{,} against the magnetic flux $\Phi$ as in Fig.\ref{Imax}.  
As can been seen from eq.(\ref{totalcurrent2}), $|J_{\rm max}|$ is to be given as a cosine curve, 
and the result of the absolute cosine curve like Fig.\ref{Imax} is one of the specific behaviors in the SQUID.   
\textcolor{black}{Here} we can see that Fig.\ref{Imax} \textcolor{black}{has a} disparity 
\textcolor{black}{ between the left and the right in the positions of the blue points}. 
The reason for that disparity is simply that our actual numerical data obtained in each Josephson junction\textcolor{black}{, as in Table.\ref{nnn},} 
is from about $-\pi/2$ to $\pi/2$ and the amount of the supercurrent in one side is fixed to a finite value. 

\section*{Acknowledgment}
I would like to thank the authors \textcolor{black}{in the} paper~\cite{Cai:2013sua}, Yong-Qiang Wang,~Rong-Gen Cai and Hai-Qing Zhang. 
Particularly I would like to thank Hai-Qing Zhang very much that \textcolor{black}{he could discuss to the end}.
\textcolor{black}{I would also like to thank Li-Fang Li for her work in the early stage of this study, 
and all the reviewers who could read the manuscript and give comments and indications.} 
Further I would like to specially thanks to James Grace in the language center of Naresuan University that he could kindly check the manuscript, 
\textcolor{black}{and I would like to thank} the warm hospitality of Tohoku University, Astronomical Institute. 
Lastly I would like to offer thanks to \textcolor{black}{the staffs} in The Institute for Fundamental Study \textcolor{black}{in Naresuan University}. 
\appendix

\section{Damping of the supercurrent in the SQUID}
\label{app:1}

In this appendix, we show that, when closing in on the center of the section 
in the superconductor part of the SQUID in Fig.\ref{FigSQUID}, the flow of the supercurrent diminishes. By this, 
we show the validity of the ${\bf v}_s=0$ in the below eq.(\ref{twopin2}).

To this purpose, taking the orthogonal coordinate system $(x,y,z)$, we assume that the section of the circuit in the superconductor part 
of the SQUID is put perpendicularly to $y$-direction and parallel with $(x,z)$ plane. Further, we direct $z$-direction parallel to the magnetic flux 
$\vec{B}\equiv \nabla \times \vec{A}$.

\textcolor{black}{Then} from the London equation 
$\displaystyle \nabla \times \vec{J}_s = - \frac{n_s e^*{}^2}{m^*} \vec{B}$ 
and \textcolor{black}{the} Maxwell equation 
$\displaystyle \nabla \times \vec{B} = \vec{J}_s$, 
we can obtain the following equation 
\begin{eqnarray}\label{nabla2vecJs}
\nabla^2 \vec{J}_s = \lambda \vec{J}_s,  
\end{eqnarray}  
where $\displaystyle \lambda \equiv \frac{n_s e^*{}^2}{m^*}$, 
and $\vec{J}_s$, $m^*$, $e^*$ and $n_s$ are the supercurrent, the mass, the electric charge and the density of the Cooper pair, respectively. 
Here it turns out that $\displaystyle \vec{J}_s=-\frac{n_s e^*{}^2}{m^*}\vec{A}$ from the London equation,  
and, in the derivation of eq.(\ref{nabla2vecJs}), we have taken a gauge fixing condition $\displaystyle \nabla \cdot \vec{A}=0$. 
Then we can obtain the solution as
\begin{eqnarray}
\vec{J}_s = e^{-\sqrt{\lambda} x},
\end{eqnarray} 
where the $x$ appearing here is the coordinate for the inside of the section of the SQUID.

Now we can see from the solution shown above, when closing in on the center of the section 
in the superconductor part of the SQUID, the flow of the supercurrent diminishes. Hence, it is reasonable to consider that the flow of the supercurrent 
vanishes at the center of the section and ${\bf v}_s=0$ as below eq.(\ref{twopin2}).

\section{Numerical results used in Figs.\ref{sine_relation} and \ref{Imax}}
\label{app:2}

We show explicitly the phase difference $\gamma_{L,R}$ defined in eq.(\ref{phasedifference}).  
These are obtained from solving the equations of motion (\ref{eom1a})-(\ref{eom5a}) in each left and right space 
one at a time by varying the value of the supercurrent $J_{L,R}$ as the inputs and taking the chemical potential as in the right figure of Fig.\ref{shape_mu}.  
Figs.\ref{sine_relation} and \ref{Imax} are plotted based on these numerical results.
Here let us notice that we measure the phases in an anticlockwise direction in Fig.\ref{FigSQUID} and 
the relation between $J_{\rm total}$ and $J_{L,R}$ are
given eq.(\ref{totalcurrent1}). 
\begin{table}[h!!!!]
\begin{center}
\begin{tabular}{|c|c|c|c|c|c|} 
 \hline 
 $ \gamma_{L,R} $ & $J_{L,R}/T_{\rm cH}{}^2$ & & $ \gamma_{L,R}  $ & $J_{L,R}/T_{\rm cH}{}^2$  \\ 
 \hline  \hline 
  -1.51614 & -1.21316   & & 0.136253 & 0.160684 \\ \hline
  -1.45013 & -1.20513   & & 0.274963 & 0.321368 \\ \hline
  -1.20955 & -1.12479   & & 0.419009 & 0.482052 \\ \hline
-0.939785 & -0.964103 & & 0.572359 & 0.642735 \\ \hline
-0.741505 & -0.803419 & & 0.741505 & 0.803419 \\ \hline
-0.572359 & -0.642735 & & 0.939785 & 0.964103 \\ \hline
-0.419009 & -0.482052 & & 1.20955   & 1.12479  \\ \hline
-0.274963 & -0.321368 & & 1.45013   & 1.20513  \\ \hline
-0.136253 & -0.160684 & & 1.51614   & 1.21316  \\ \hline
             0 &              0 & &               &              \\ \hline
\end{tabular} 
\caption{
\textcolor{black}{
The Chebyshev Grid is taken for the all as $(n_z,n_\chi)=(20,35)$ except for for the ones of the $J_{L,R}/T_{\rm cH}{}^2=\pm 1.21326$. 
The Chebyshev Grid taken for the ones of the $J_{L,R}/T_{\rm cH}{}^2=\pm 1.21326$ is $(n_z,n_\chi)=(23,43)$.}  
Here $n_z$ and $n_\chi$ mean the number or the grid in $z$- and $\chi$-directions, respectively. 
The critical temperature for the superconductor/normal metal transition $T_{\rm cH}$ is defined as $T_{\rm cH}= c \, \mu_{\rm H}$ with $c \equiv 0.0588$~\cite{Horowitz:2011dz}. 
}
\label{nnn}
\end{center}
\end{table}


\begin{thebibliography}{99}


\bibitem{Maldacena:1997re} 
  J.~M.~Maldacena,
  ``The Large N limit of superconformal field theories and supergravity,''
  Adv.\ Theor.\ Math.\ Phys.\  {\bf 2}, 231 (1998)
  [hep-th/9711200].
  

\bibitem{Gubser:1998bc}
  S.~S.~Gubser, I.~R.~Klebanov and A.~M.~Polyakov,
  ``Gauge theory correlators from non-critical string theory,''
  Phys.\ Lett.\ B {\bf 428}, 105 (1998)
  [arXiv:hep-th/9802109].
 
\bibitem{Witten:1998qj}
  E.~Witten,
  ``Anti-de Sitter space and holography,''
  Adv.\ Theor.\ Math.\ Phys.\  {\bf 2}, 253 (1998)
  [arXiv:hep-th/9802150].
  
 
\bibitem{Gubser:2008px} 
  S.~S.~Gubser,
  ``Breaking an Abelian gauge symmetry near a black hole horizon,''
  Phys.\ Rev.\ D {\bf 78}, 065034 (2008)
  [arXiv:0801.2977 [hep-th]].

\bibitem{Hartnoll:2008vx} 
  S.~A.~Hartnoll, C.~P.~Herzog and G.~T.~Horowitz,
  ``Building a Holographic Superconductor,''
  Phys.\ Rev.\ Lett.\  {\bf 101}, 031601 (2008)
  [arXiv:0803.3295 [hep-th]].
  
\bibitem{Hartnoll:2008kx} 
  S.~A.~Hartnoll, C.~P.~Herzog and G.~T.~Horowitz,
  ``Holographic Superconductors,''
  JHEP {\bf 0812}, 015 (2008)
  [arXiv:0810.1563 [hep-th]].
  
   
\bibitem{Lee:2008xf} 
  S.~-S.~Lee,
  \textcolor{black}{``A Non-Fermi Liquid from a Charged Black Hole: A Critical Fermi Ball,''}
  Phys.\ Rev.\ D {\bf 79}, 086006 (2009)
  [arXiv:0809.3402 [hep-th]].
   
\bibitem{Liu:2009dm} 
  H.~Liu, J.~McGreevy and D.~Vegh,
  ``Non-Fermi liquids from holography,''
  Phys.\ Rev.\ D {\bf 83}, 065029 (2011)
  [arXiv:0903.2477 [hep-th]].
  
\bibitem{Cubrovic:2009ye} 
  M.~Cubrovic, J.~Zaanen and K.~Schalm,
  ``String Theory, Quantum Phase Transitions and the Emergent Fermi-Liquid,''
  Science {\bf 325}, 439 (2009)
  [arXiv:0904.1993 [hep-th]].


\bibitem{Nishioka:2009zj} 
  T.~Nishioka, S.~Ryu and T.~Takayanagi,
  ``Holographic Superconductor/Insulator Transition at Zero Temperature,''
  JHEP {\bf 1003}, 131 (2010)
  [arXiv:0911.0962 [hep-th]].

  
\bibitem{Cai:2015cya} 
  \textcolor{black}{R.~G.~Cai, L.~Li, L.~F.~Li and R.~Q.~Yang,
  ``Introduction to Holographic Superconductor Models,''
  arXiv:1502.00437 [hep-th].}
  
  
\bibitem{Josephson:1962zz}
  B.~D.~Josephson,
  {\em  Possible new effects in superconductive tunnelling},
  Phys.\ Lett.\  {\bf 1}, 251 (1962).
  
  
\bibitem{Horowitz:2011dz} 
  G.~T.~Horowitz, J.~E.~Santos and B.~Way,
  {\em A Holographic Josephson Junction},
  Phys.\ Rev.\ Lett.\  {\bf 106}, 221601 (2011)  [arXiv:1101.3326 [hep-th]].  
   

   
    
\bibitem{Wang:2011rva} 
  Y.~-Q.~Wang, Y.~-X.~Liu and Z.~-H.~Zhao,
  ``Holographic Josephson Junction in 3+1 dimensions,''
  arXiv:1104.4303 [hep-th]. 
  
\bibitem{Siani:2011uj} 
  M.~Siani,
  ``On inhomogeneous holographic superconductors,''
  arXiv:1104.4463 [hep-th].
  
\bibitem{Wang:2011ri} 
  Y.~-Q.~Wang, Y.~-X.~Liu and Z.~-H.~Zhao,
  ``Holographic p-wave Josephson junction,''
  arXiv:1109.4426 [hep-th].
  
\bibitem{Wang:2012yj} 
Y.~-Q.~Wang, Y.~-X.~Liu, R.~-G.~Cai, S.~Takeuchi and H.~-Q.~Zhang,
  ``Holographic SIS Josephson Junction,''
  JHEP {\bf 1209}, 058 (2012)
  [arXiv:1205.4406 [hep-th]].
  

\bibitem{Kiritsis:2011zq} 
  E.~Kiritsis and V.~Niarchos,
  ``Josephson Junctions and AdS/CFT Networks,''
  JHEP {\bf 1107}, 112 (2011)
  [Erratum-ibid.\  {\bf 1110}, 095 (2011)]
  [arXiv:1105.6100 [hep-th]].

  
\bibitem{Li:2014xia} 
  H.~F.~Li, L.~Li, Y.~Q.~Wang and H.~Q.~Zhang,
  ``Non-relativistic Josephson Junction from Holography,''
  JHEP {\bf 1412}, 099 (2014)
  [arXiv:1410.5578 [hep-th]].
  


 \bibitem{Nakano:2008xc} 
  E.~Nakano and W.~-Y.~Wen,
  ``Critical magnetic field in a holographic superconductor,''
  Phys.\ Rev.\ D {\bf 78}, 046004 (2008)
  [arXiv:0804.3180 [hep-th]]. 

\bibitem{Albash:2008eh}  
  T.~Albash and C.~V.~Johnson,
  ``A Holographic Superconductor in an External Magnetic Field,''
  JHEP {\bf 0809}, 121 (2008)
  [arXiv:0804.3466 [hep-th]]. 

\bibitem{Hartnoll:2007ai} 
  S.~A.~Hartnoll and P.~Kovtun,
  ``Hall conductivity from dyonic black holes,''
  Phys.\ Rev.\ D {\bf 76}, 066001 (2007)
  [arXiv:0704.1160 [hep-th]]. 

  
\bibitem{Domenech:2010nf}  
  O.~Domenech, M.~Montull, A.~Pomarol, A.~Salvio and P.~J.~Silva,
  ``Emergent Gauge Fields in Holographic Superconductors,''
  JHEP {\bf 1008}, 033 (2010)
  [arXiv:1005.1776 [hep-th]].


\bibitem{Montull:2011im} 
  M.~Montull, O.~Pujolas, A.~Salvio and P.~J.~Silva,
  ``Flux Periodicities and Quantum Hair on Holographic Superconductors,''
  Phys.\ Rev.\ Lett.\  {\bf 107}, 181601 (2011)
  [arXiv:1105.5392 [hep-th]]. 


\bibitem{Salvio:2013ja} 
\textcolor{black}{
  A.~Salvio, 
  ``Superconductivity, Superfluidity and Holography,''
  J.\ Phys.\ Conf.\ Ser.\  {\bf 442}, 012040 (2013)
  [arXiv:1301.0201]. }

\bibitem{Montull:2012fy} 
  M.~Montull, O.~Pujolas, A.~Salvio and P.~J.~Silva,
  ``Magnetic Response in the Holographic Insulator/Superconductor Transition,''
  JHEP {\bf 1204}, 135 (2012)
  [arXiv:1202.0006 [hep-th]].
  
\bibitem{Cai:2012vk} 
  R.~-G.~Cai, L.~Li, L.~-F.~Li, H.~-Q.~Zhang and Y.~-L.~Zhang,
  ``Wilson Line Response of Holographic Superconductors in Gauss-Bonnet Gravity,''
  Phys.\ Rev.\ D {\bf 87}, 026002 (2013)
  [arXiv:1209.5049 [hep-th]].
  

  \bibitem{JosephsonMaterial}
  Pasi L$\ddot{\rm a}$hteenm$\ddot{\rm a}$ki, G. S. Paraoanu, Juha Hassel, Pertti J. Hakonen,
  ``Dynamical Casimir effect in a Josephson metamaterial,''
  Proc. Natl. Acad. Sci. U.S.A. {\bf 110}, 4234 (2013)   [arXiv:1111.5608].


\bibitem{Cai:2013sua} 
  R.~-G.~Cai, Y.~-Q.~Wang and H.~-Q.~Zhang,
 ``A holographic model of SQUID,''
  arXiv:1308.5088 [hep-th].
  
 
\bibitem{Horowitz:1998ha} 
  G.~T.~Horowitz and R.~C.~Myers,
  ``The AdS / CFT correspondence and a new positive energy conjecture for general relativity,''
  Phys.\ Rev.\ D {\bf 59}, 026005 (1998)
  [hep-th/9808079].
  


\bibitem{Nakamura:2007nk} 
\textcolor{black}{
  S.~Nakamura,
  ``Comments on Chemical Potentials in AdS/CFT,''
  Prog.\ Theor.\ Phys.\  {\bf 119}, 839 (2008)
  [arXiv:0711.1601 [hep-th]].}
  
 
  \bibitem{trefethn}
Lloyd N. Trefethen, {\it Spectral Methods in MATLAB, SIAM, Philadelphia, 2000}.


\end{thebibliography}
\end{document}